\documentstyle[prl,aps,epsf]{revtex}
\begin{document}
  
\draft
\twocolumn[\hsize\textwidth\columnwidth\hsize\csname@twocolumnfalse\endcsname

\title{Defects, order, and hysteresis  in driven charge-density waves}
 
\author{Mikko Karttunen$^{1}$, Mikko Haataja$^{1}$, 
        K. R. Elder$^{2}$, and Martin Grant$^{1}$}

\address{$^1$Department of Physics and Centre for the Physics of
         Materials, McGill University, 3600 rue University, 
         Montr\'eal (Qu\'ebec), Canada H3A 2T8}

\address{$^2$Department of Physics, Oakland University, 
         Rochester, MI, 48309-4487}

\date{\today}

\maketitle

\begin{abstract}
We model driven two-dimensional charge-density waves in random media
via a modified Swift-Hohenberg equation, which includes both amplitude
and phase fluctuations of the condensate. As the driving force is
increased, we find that the defect density first increases and then 
decreases.  Furthermore, we find switching phenomena, due to the 
formation of channels of dislocations.  These results are in 
qualitative accord with recent dynamical x-ray scattering experiments 
by Ringland {\it et al.\/}\ and transport experiments by 
Lemay {\it et al.} 
\end{abstract}


\vskip1pc]
\narrowtext

The effect of quenched disorder on a periodic medium has been the
subject of intense study during the past decade. The rich phenomena
displayed by these systems has raised intriguing questions 
\cite{ringland99a,lemay99a,preringland99b}.  These
include the nature of the depinning transition, the effect of disorder
on structural properties, and the possibility of the existence of
nonequilibrium analogs of solids and liquids. Examples of such systems
include charge-density waves (CDW) \cite{gruner88a}, flux line lattices
in type-II superconductors\cite{dsfisher91a,koshelev94a}, and magnetic
bubble arrays \cite{hu95a}.  A common effect of quenched disorder in
equilibrium systems displaying periodicity is to reduce or destroy that
periodicity, by pinning the system in a disordered state.
However, a sufficiently large driving force in a nonequilibrium system
can act to depin the system, thereby reducing  the effect of quenched
disorder to that of annealed disorder, analogous to thermal noise.

A CDW is a periodic modulation of the electron density, of wavenumber
$Q$, which results from the electron-phonon interaction
\cite{bpeierls55a}.  The local excess density of the conduction
electrons is given by $ \rho(\vec{x},t)=\rho_{c} (\vec{x},t) \cos
(\vec{Q} \cdot \vec{x}+ \phi(\vec{x},t))$, where $Q \equiv 2k_F$, $k_F$
is the Fermi wavenumber, $\rho_c$ is the CDW amplitude, and $\phi$ is
the phase. It is convenient to introduce a scalar order parameter
$\psi(\vec{x},t)= \mbox{Re}[ e^{i \vec{Q} \cdot \vec{x}}
\rho_c(\vec{x},t)e^{i \phi(\vec{x},t)}]$.  

An important question concerns the role of amplitude fluctuations 
of the condensate close to the depinning transition 
\cite{coppersmith90a,coppersmith91a,coppersmith91b,balents95a,tang87a}. 
A common approximation is to neglect amplitude fluctuations,
leading to a phase-only description
\cite{fukuyama76a,fukuyama78a,lee79a,RAMAKRISHNA92}. However, this
approximation is not always justified, as demonstrated
by Coppersmith and Millis 
\cite{coppersmith90a,coppersmith91a,coppersmith91b} 
for a model of a driven CDW, and as argued 
by Balents and Fisher \cite{balents95a}. Moreover,
recent theoretical studies in 2D indicate that
dislocations proliferate in the presence of
quenched disorder even in the absence of
an external driving force \cite{zeng99a}. Therefore,
the role of amplitude fluctuations close to the
depinning transition warrants a closer 
examination.
 
To this end, we propose a model for driven
CDWs, and show that dislocations indeed
proliferate close to the depinning transition. 
This behavior has profound implications on the
structure of the CDW and its transport properties.
In particular, we show how the dynamical generation 
of dislocations can make the system
{\it more disordered\/} above the 
depinning threshold. This result is in agreement
with recent x-ray scattering experiments by
Ringland {\it et al.\/}\ \cite{ringland99a}
who observed an increasing correlation length 
upon letting the CDW relax from the sliding state
to the pinned state. This behavior, generic in
our model, has not been seen in any of the previously 
studied theoretical models 
\cite{sibani90a,middleton92a,chen96a}.
Furthermore, our results suggest that upon further
increasing the drive, the system becomes more
ordered. This is in accord with very recent experimental results
by Ringland {\it et al.\/}\ \cite{preringland99b}.
In particular, we predict that the 
correlation length should increase exponentially.
These results are consistent with 
the picture of Balents and Fisher \cite{balents95a} 
of a dynamical phase transition 
into a temporally 
periodic ``moving solid'' phase.
Finally, we show that the behavior of the
dislocation density can be hysteretic upon increasing
and decreasing the drive.
Indeed, we find that the hysteresis is associated 
with the activated formation of 
channels of dislocations.
Based on these results we argue
that the current in the model should display 
``switching'' (i.e., hysteresis). This is in
qualitative agreement with recent transport
experiments by Lemay {\it et al.\/}\ \cite{lemay99a}.

We propose a phenomenological equation
of motion, related to the Swift-Hohenberg equation 
\cite{swift77a}, which permits 
periodic solutions to occur only in the $x$ direction,

\begin{eqnarray}
\frac{\partial  \psi}{\partial t} + 
E \frac{\partial  \psi}{\partial x}
 & = \left[ \epsilon - \left(q_c^2+ 
\frac{\partial^2}{\partial x^2}\right)^2 
+ 4 \, q_c^2 \,\frac{\partial^2}{\partial y^2} 
\right] \psi \nonumber \\ & - \psi^3 
 + \sum_i V_{imp}^i,
\label{eq:msh}
\end{eqnarray}
where $\psi =\psi(\vec{x},t) $ denotes the local condensate density, 
and $\epsilon \equiv |T-T_p|/T_p$, with $T_p$ denoting
the critical temperature of the Peierls transition. 
$E$ denotes the external drive, and $V_{imp}^i$ is the localized
impurity pinning. $q_c$ stands for the wavenumber of the
periodic modulation which is set to unity hereafter.
Impurities are
characterized by a concentration $c$ and quenched Gaussian correlations
$\langle V_{imp}^i \rangle=0$ and $\langle V_{imp}^i V_{imp}^j \rangle
= 2 V_0 \delta_{ij}$. 
The use of both impurity strength $V_0$ and impurity concentration $c$
to describe impurities provides a convenient way to tune between weak and
strong pinning regimes, as described below.
We note
that the dynamics of $\psi$, in the absence
of external drive, follow from the Lyapunov functional:
\begin{eqnarray} 
F = &\int d\vec{x} \Bigl\{ -\psi \Bigl[
\epsilon- \Bigl(q_c^2+ \frac{\partial^2} {\partial x^2}\Bigr)^2
 + 4 \, q_c^2 \,\frac{\partial^2}{\partial y^2} \Bigr] \frac{\psi}{2}
\nonumber \\ 
          &  + \psi^4/4 + \sum_i \psi V_{imp}^i \Bigr\}.  
\end{eqnarray}
That is, $\partial \psi/\partial t = -\delta F / \delta \psi$.  It is
straightforward to include the effects of thermal noise
\cite{karttunen99c}.

This equation naturally produces periodic structures, permitting the
study of local fluctuations in the wavelength,
as well as order and defects, as described by the phase and
amplitude variations of the charge-density wave.
In particular, a
second-order derivative is sufficient in the $y$-direction since no
periodic ordering occurs in the that direction, while 
both second and fourth-order derivatives are necessary in the
$x$-direction to produce a periodic pattern, and to suppress spatial
gradients
on very small length scales.  Higher-order derivatives in the $x$
or $y$ directions may alter the local structure of the defects, but will
not qualitatively change the long-range interaction between defects.
In the limit of 
small $\epsilon$, it is equivalent to the standard Ginzburg-Landau approach
\cite{gruner88a,fukuyama76a,fukuyama78a,lee79a}.
The equivalence can be shown via a multiple-scales
expansion \cite{karttunen99c} using the slow variables 
$\epsilon t$, $\epsilon^{1/2} x$, and $\epsilon^{1/2} y$, and the
ansatz, $\psi = 
\epsilon^{1/2} \psi_0 + \epsilon \psi_1 + \epsilon^{3/2} \psi_2 +
\ldots$.  To leading nontrivial order, we obtain 
the equation of motion 
for the complex amplitude $A$ for the most unstable mode
${\partial A}/{\partial t} = 4 ({\partial^2}/{\partial x^2} + 
{\partial^2}/{\partial y^2}) A + A - 3|A|^2 A$,
which is the time-dependent Ginzburg-Landau equation of motion for
a complex order parameter, which includes both phase and amplitude
fluctuations.  The effects
of driving force and pinning potential can be straightforwardly
included.  This, together with an explicit mapping of our approach to 
that of others, will be given in a future paper \cite{karttunen99c}.
The advantage of our approach is the natural interplay between local
variations in wavelength during the dynamical process, which is
difficult to isolate in the Ginzburg-Landau equation. 
Away from threshold, $\epsilon \sim 1$, other unstable modes can become
important.  We expect that, since 
our equation captures the correct symmetries
and essential physics of the driven charge density waves, that it will
remain a valid description for large $\epsilon$.
Indeed, $q_c$ is only weakly dependent on $\epsilon$ \cite{pomeau79a}.  
Also, our numerical results and tests for varying driving forces and
$\epsilon$ show no qualitative dependence on 
the size of $\epsilon$.

The nature of CDW pinning is determined by comparing the energy gain
the wave experiences by adjusting to the local impurities,
to the increased elastic energy due that adjustment \cite{GRUNER}.  
The former energy is  $V_0 c \epsilon^{1/2} $, where
$\epsilon^{1/2}$ gives the amplitude of the CDW.  The latter is
$\epsilon c \ell^{-2}$, where $\epsilon$ gives the elastic force
constant, and $\ell \sim c^{-1/d}$ is the distance between impurities
in $d$ dimensions.  In two dimensions, the
ratio of these gives a dimensionless
parameter $\kappa = V_0/(c\epsilon^{1/2})$.  Strong pinning corresponds
to large $\kappa$, weak pinning to small $\kappa$.  Herein we
conveniently tune between regimes by varying $c$, so that strong pinning
corresponds to small $c$.

\begin{figure}[!]
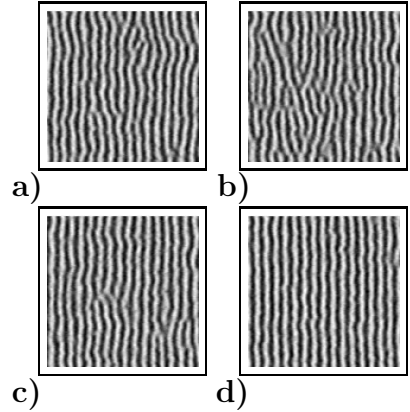

\hspace*{0.5cm}\fbox{\epsfxsize=2cm
\epsfbox{con_0.0.epsi}} 
\hspace*{0.2cm}
\fbox{\epsfxsize=2cm
\epsfbox{con_0.05.epsi}} \\
\hspace*{0.5cm}{\large \bf a) \hspace*{2cm} b)}\\
\hspace*{0.85cm}\fbox{\epsfxsize=2cm
\epsfbox{con_0.1.epsi}} 
\hspace*{0.2cm}
\fbox{\epsfxsize=2cm
\epsfbox{con_0.2.epsi}} \\
\hspace*{0.5cm}{\large \bf c) \hspace*{2cm} d)}\\ 
\caption{(a)-(d): Typical configurations. In (a), the external drive
$E=0$, $V_0 = 0.4$ and $c=0.8$. 
In (b), $E = 0.05$. In (c), $E=0.1$. In (d), $E=0.2$. Only one
quarter of the original system size is shown for clarity.}
\end{figure} 

We studied the discretized version of Eq.~(\ref{eq:msh}) 
on a $d=2$ square
lattice of size $L=256$, where the discretized elements of time and space
were $dt=0.05$ and $dx=0.78$, respectively.
Gradients were evaluated using standard finite differencing.
We verified that making $dt$ and $dx$ smaller did not
change our results.  
The control parameter, $\epsilon$, was varied 
between $0.1$ and $1.2$, 
the pinning strength $V_0 = 0.1-1.5$, the
concentration of pinning sites $c = 0-1$, and the driving force $E=
0-0.2$.  Results involve averages over at least 
10 independent representations of the disorder.
Typical configurations of the system are shown in Fig.~1 with
$c=0.8$ and $V_0=0.4$.
In the relaxed state ($E=0$), Fig.~1a,
some defects are visible.  As the driving force is
increased past the depinning transition (Fig.~1b with $E = 0.05$) the
number of defects $n_D$ increases \cite{Ec}.  
However, for a still larger driving force,
defects begin to disappear (Fig.~1c with $E = 0.1$)
and, eventually, the system becomes more ordered (Fig.~1d
where $E= 0.2$).

We quantify this observation by examining the 
spherically averaged structure factor $S(q)$, defined by
$S(q) = \langle |\hat{\psi}(\vec{q}\,')|^2 \rangle_{|\vec{q}\,' | = q}$,
where $\hat{\psi}$ denotes the Fourier
transform of $\psi$
and $\langle \ldots \rangle_{|\vec{q}\,' | = q}$
denotes an average over all configurations and 
Fourier modes with $|\vec{q}\, '| = q$. 
A typical structure factor is shown in Fig.~2. for a 
relaxed state $E=0$, and for increasing driving force in the
sliding state where $E> E_c$ ($E= 0.025$ and $0.05$). 
As the drive is turned off, the peak of the
structure factor increases and, correspondingly, the half
width at half maximum decreases. This implies
that the correlation length increases upon
switching off the drive, in qualitative
agreement with recent dynamical x-ray scattering
experiments by Ringland {\it et al.\/} 
\cite{ringland99a}. Furthermore, upon monotonically
increasing the drive we observe a sharpening of the
peak, due to decreasing $n_D$. 
This surprising nonmonotonic behavior
is in good qualitative accord 
with recent x-ray scattering experiments by
Ringland {\it et al.\/}\ \cite{ringland99a,preringland99b}.

\begin{figure}[!]
\hspace*{0.25cm} \epsfxsize=7cm
\epsfbox{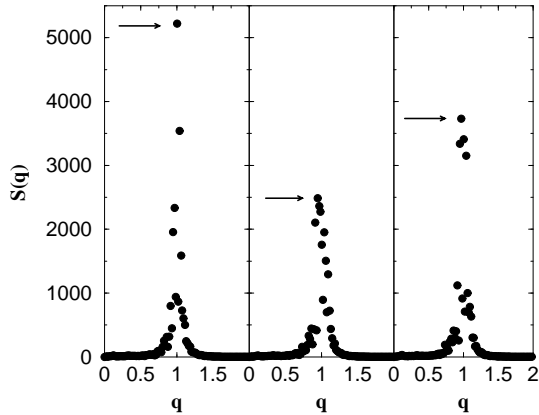}
\vspace*{0.5cm}
\caption{Structure factor $S(q)$ for, from left to right, $E= 0$, $E=
0.075$ and $E= 0.09$. $V_0=0.5$ and $c=0.8$. 
}
\end{figure}

To further quantify this nonmonotonic dependence, we estimated 
numerically the number of dislocations $n_D$ in the
system in the following manner. First we determined the number of 
areas $n_A$ in the system where the local {\it transverse} gradient 
$|\nabla_y \psi| \geq 0.35$. We have checked that 
$n_A = A\cdot  n_D + B$, where $A \approx 10$  is the contribution
from an isolated defect, and  $B$  is a measure of
the roughness of the CDW in the absence of defects, which is dependent
on $c$.  In Fig.~3 we show the number of dislocations $n_D$ 
for $c=0.4$, corresponding
to the strong pinning regime; the weak pinning result ($c=0.8$)
is shown in the inset. As Fig.~1 indicated, there is an initial
increase in $n_D$, up to a maximum appearing close to
the depinning transition, followed by a subsequent decay
for $E > E_c$.

The behavior of $n_D$ and the structure factor
can be understood using a simple argument.
In the absence of thermal fluctuations, the generation of 
dislocations must depend 
on the number density of pinned sites $n_p$ and on 
the local strain rate. In addition, 
annihilation of dislocations depends on 
the mobility $D$ and $n_D$. In mean-field theory, 
the rates for generation
and annihilation of defects are given by 
$R_G \sim n_p \cdot v$ and  $R_A \sim D \cdot n_D$.
The velocity $v$ of the CDW enters into $R_G$ 
since the local strain rate is proportional to 
the local velocity. Using the above expressions, 
the mean-field steady-state number 
of dislocations is simply
$n_D \sim n_p \cdot v$. Scaling arguments imply
\cite{karttunen99c} 
$n_D \sim (E-E_c)^{\theta - \nu}$ as $E \rightarrow E_c$
and $n_D \sim \exp -E^2$
as $E \gg E_c$, where $\theta$ and
$\nu$ denote the velocity and correlation length
exponents, respectively.
Indeed, our data is in good qualitative agreement with these
arguments, although our data do not permit a detailed test.

\begin{figure}[!]
\hspace*{0.25cm} \epsfxsize=7cm
\epsfbox{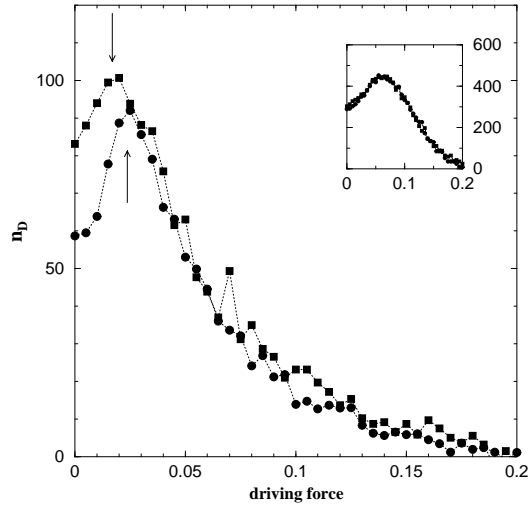}
\vspace*{0.5cm}
\caption{Number of dislocations $n_D$ as a
function of the driving force $E$ for strong pinning. 
Pinning strength
$V_0 = 0.5$ and impurity concentration $c=0.4$. Circles:
increasing drive. Squares: decreasing drive. Inset: $n_D$ for
weak pinning ($c=0.8$).}
\end{figure}

\begin{figure}[!]
\hspace*{0.5cm}\fbox{\epsfxsize=2cm
\epsfbox{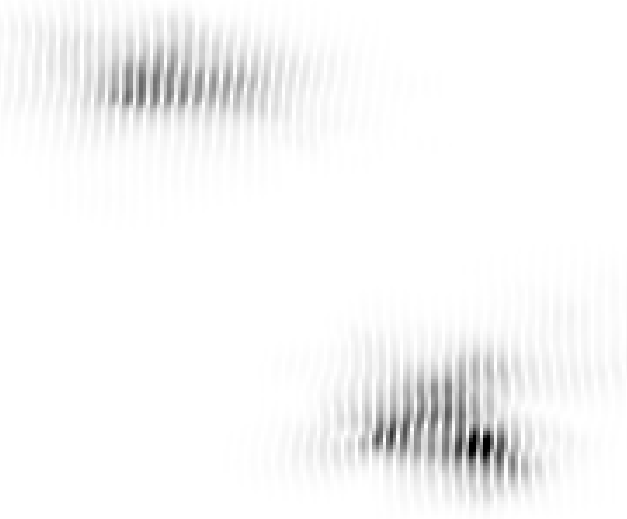}} 
\hspace*{0.2cm}
\fbox{\epsfxsize=2cm
\epsfbox{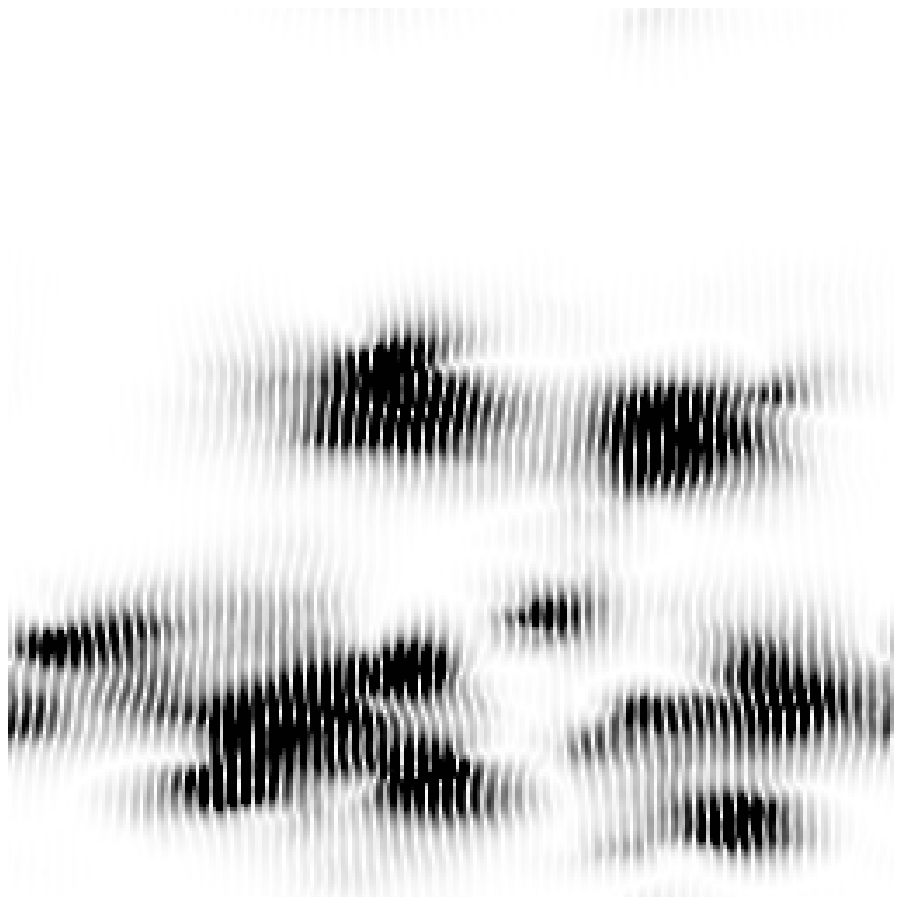}}
\hspace*{0.2cm}
\fbox{\epsfxsize=2cm
\epsfbox{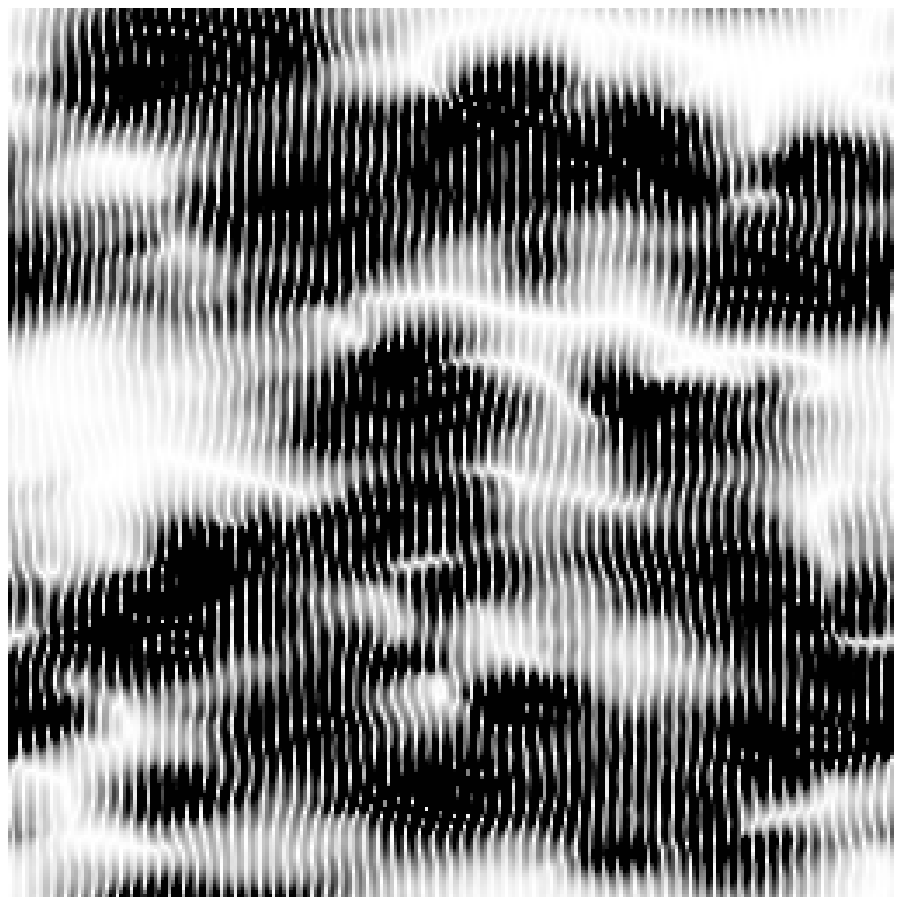}}
 \\
\hspace*{0.5cm}{\large \bf a) \hspace*{2cm} b) \hspace*{2cm} c)}\\
\caption{The behavior of the system near the 
depinning threshold in the strong pinning regime. Each of the snapshots 
is a difference between two configurations $500$ time steps $dt$
apart (25 time units since $dt= 0.05$). 
The black areas indicate where
the changes occur, i.e., which areas are moving 
The grey scale is the same in all the pictures. 
$V_0=0.5$, $c=0.4$. (a) at $E=0.020$, (b)  $E=0.025$, and 
(c) $E=0.030$.
}
\end{figure}

In the strong pinning regime, additional behavior is evident:
$n_D$ shows significant {\it hysteresis\/} as the drive is
first increased and then decreased. In 
the weak pinning limit, no significant hysteresis 
is seen. The origin of this is evident in Fig.~4.
We compare two configurations
of the same system at different time steps, around the depinning
threshold.  The configurations were taken 500 time steps $dt$
apart (25 time units since $dt= 0.05$), and then
they were deducted from each other.  The
darker areas indicate where the changes are the largest.
Below the threshold, in the strong pinning regime,
there is little movement in the system.
As the driving force is increased to the transition, 
a conducting channel appears.  As 
the driving force is increased further, the system depins coherently.
In other words, depinning in the strong pinning regime is mediated by
activated channels \cite{balents96a}.

In contrast, in the weak pinning regime (not shown) domains form 
cooperatively, even before depinning takes place, and the transition
takes place collectively, with insignificant 
apparent activation.  Unfortunately, 
in our two-dimensional simulations, we cannot
determine if the qualitative transition in behavior
between weak and strong pinning
is sharp, or governed by crossover.
The qualitative difference is easy to understand.
In the strong-pinning regime, impurities are far apart, so
channels, where regions of current flow,
``nucleate'' between impurities.
In the weak pinning regime, where there are more
impurities, there is insufficient room for channels to form, and,
instead, a cooperative and collective effort of the entire system is
required to depin the system.
This bistable behavior in the strong pinning regime
will manifest itself in the
current-voltage characteristics of
the sample.  This is known as ``switching'', and 
is a generic feature of low-temperature CDWs
\cite{lemay99a,zettl82a,hall84a,hall88a}.
This behavior, which follows naturally from our approach,
is in qualitative 
agreement with recent transport measurements of Lemay {\it et al.} 
\cite{lemay99a}.

In conclusion, we have introduced a novel approach to
study the of driven charge-density waves. 
We have demonstrated that, above the global depinning
threshold, phase slips appear and lead to an even
more disordered structure for the CDW, as seen experimentally
\cite{ringland99a}.
However, far away
from threshold the system becomes more ordered, as observed very
recently \cite{preringland99b}.  These latter results are in 
agreement with the picture of Balents and Fisher
\cite{balents95a}. 
Finally our model gives rise to a simple and generic mechanism for
switching in the strong pinning regime, through the
activated appearance of channels,
which qualitatively
explains recent experimental results \cite{lemay99a}.

We thank the authors of Refs.\ \cite{ringland99a,lemay99a,preringland99b}, 
particularly Joel Brock, for
useful discussions and for communicating their work to us prior to
publication.  We thank Mark Sutton for useful discussions.
This work has
been supported by the Academy of Finland (MK \& MH), 
the Finnish Cultural Foundation (MK),
the Finnish Academy of Science and Letters (MK), 
Research Corporation grant CC4787 (KRE),
the Natural Sciences and
Engineering Council of Canada (MG), {\it 
le Fonds pour la Formation de
Chercheurs et l'Aide \'a la Recherche du Qu\'ebec\/} (MG).


\begin{references}
\bibitem{Ec} 
The position of the depinning transition $E_c$ 
was determined by measuring 
where the average current $I$, estimated by $I\propto |v_x|/ L^2 $ 
becomes nonzero.  
The depinning transition is somewhat to the left of the peak in $n_D$
in Fig.~3.
This measure of $I$ does not distinguish 
dislocations from periodic waves, and so does not give
the current-voltage characteristics for $E > E_c$.
In a future long paper \cite{karttunen99c}, we will discuss the
relationship of $E_c$ to the threshold field in, for example, the
Fukuyama-Lee-Rice model.
\end{references}
\end{document}